# Residual Information of Previous Decision Affects Evidence Accumulation in Current Decision

**Farzaneh Olianezhad**[a, c], **Maryam Tohidi-Moghaddam**[b, c], **Sajjad Zabbah**[c], **Reza Ebrahimpour**[b, c*]

[a] Faculty of Electrical Engineering, Shahid Rajaee Teacher Training University, P.O. Box: 16785-136, Tehran, Iran.

[b] Faculty of Computer Engineering, Shahid Rajaee Teacher Training University, P.O. Box: 16785-163, Tehran, Iran.

[c] School of Cognitive Sciences, Institute for Research in Fundamental Sciences (IPM), P.O. Box: 19395-5746, Tehran, Iran.

**A B S T R A C T**

Bias in perceptual decisions comes to pass when the advance knowledge colludes with the current sensory evidence in support of the final choice. The literature on decision making suggests two main hypotheses to account for this kind of bias: internal bias signals are derived from (a) the residual of motor response-related signals, and (b) the sensory information residues of the decisions that we made in the past. Beside these hypotheses, a credible hypothesis proposed by this study to explain the cause of decision biasing, suggests that the decision-related neuron can make use of the residual information of the previous decision for the current decision. We demonstrate the validity of this assumption, first by performing behavioral experiments based on the two-alternative forced-choice (TAFC) discrimination of motion direction paradigms and then, we modified the pure drift-diffusion model (DDM) based on accumulation to the bound mechanism to account for the sequential effect. In both cases, the trace of the previous trial influences the current decision. Results indicate that the probability of being correct in a current decision increases if it is in line with the previously made decision. Also, the model that keeps the previous decision information provides a better fit to the behavioral data. Our findings suggest that the state of a decision variable which is represented in the activity of decision-related neurons after crossing the bound (in the previous decision) can accumulate with the decision variable for the current decision in consecutive trials.

---

∗ Correspondence to: R. Ebrahimpour, Department of Computer Engineering, Shahid Rajaee Teacher Training University, P.O. Box: 16785-163, Tel/Fax Number: +982122970117, Tehran, Iran. E-mail addresses: f.olianezhad@ipm.ir (F. Olianezhad), m.tohidi@ipm.ir (M. Tohidi-Moghaddam), s.zabbah@ipm.ir (S. Zabbah), rebrahimpour@srttu.edu (R. Ebrahimpour).

## 1. INTRODUCTION

Perceptual decisions and their outcomes can be related to each other as a sequence (Akaishi, Umeda, Nagase, & Sakai, 2014; Bornstein et al., 2017; Glimcher, 2003; Gold, Law, Connolly, & Bennur, 2008; Goldfarb, Wong-Lin, Schwemmer, Leonard, & Holmes, 2012; Hanks, Mazurek, Kiani, Hopp, & Shadlen, 2011; Miller, Shenhav, & Ludvig, 2017; Mulder, Wagenmakers, Ratcliff, Boekel, & Forstmann, 2012; Purcell & Kiani, 2016; Remington, 1969). This ability to merge the advance knowledge about choice alternatives with current evidence to make an appropriate decision is a hallmark of higher brain function (Churchland, Kiani, & Shadlen, 2008; Cook & Maunsell, 2002; Gold & Shadlen, 2007; Heitz & Schall, 2012; Kiani, Hanks, & Shadlen, 2008; Ratcliff, Hasegawa, Hasegawa, Smith, & Segraves, 2007; Roitman & Shadlen, 2002). Neural activities in brain areas involved in decision making process seem to contain the history of previous decisions (Akaishi et al., 2014; Basten, Biele, Heekeren, & Fiebach, 2010; Boettiger et al., 2007; Fleming, Thomas, & Dolan, 2010; Fleming, Whiteley, Hulme, Sahani, & Dolan, 2010; Forstmann, Brown, Dutilh, Neumann, & Wagenmakers, 2010; Mulder et al., 2012; Philiastides, Biele, & Heekeren, 2010; Preuschhof, Schubert, Villringer, & Heekeren, 2010; Scheibe, Ullsperger, Sommer, & Heekeren, 2010; Serences, 2008; Summerfield & Koechlin, 2008, 2010) for a specific period and do not return to the initial baseline at the time of decision making. Recent findings suggest that there is a preference in humans to make decisions similar to previous ones, especially in an ambiguous stimulus (Akaishi et al., 2014; Brehm, 1956; Izuma & Murayama, 2013). This interaction between the history of choices and sensory context, respectively called internal and external signals, is thought to cause the biased decisions about the sensory events (Akaishi et al., 2014; Albright, 2012; Awh, Belopolsky, & Theeuwes, 2012; Carnevale, de Lafuente, Romo, & Parga, 2012).

The mechanism of decision bias as one of the most pervasive biases across many domains of cognitive science, however, remains obscure (Glimcher, 2003; Hanks & Summerfield, 2017; Kim, Kabir, & Gold, 2017; Lauwereyns, 2010; Summerfield & Koechlin, 2010; White & Poldrack, 2014). Two main hypotheses have been proposed to explain the reasons of this bias, although to date, none of them have been adequately supported. According to the first view, the residual of the sensory information of the previous stimulus causes internal bias signals (Akaishi et al., 2014; Albright, 2012; Carnevale et al., 2012; Pearson & Brascamp, 2008). Therefore, a strong sensory signal in the previous trial raises the neural response in the brain sensory areas and the current decision is expected to be made under a larger bias (Akaishi et al., 2014). In the alternative view, the residual of motor response-related signals causes internal bias signals (Gold et al., 2008; Marcos et al., 2013); however, contrary to the first impression, the strength of the sensory signal in the previous trial does not seem to affect the decision-biasing. Akaishi et al. also suggest that in the absence of feedback, this bias is a mechanism to update the likelihood of a choice to be made (Akaishi et al., 2014).

Over and above the mentioned assumptions, the following hypothesis is proposed in this study as plausible: The residual evidence of the previous decision in the decision-related neuron can be informative for the current decision. We tested the validity of this claim using behavioral experiments based on the two-alternative forced-choice (TAFC) discrimination of motion direction and modeling approach. We revealed that firstly, the probability of being correct in the current decision increases if it is in line with the previous decision, showing a trace from the previous trial on the current one. Secondly, this effect is independent of the correctness of the previous decision and the feedback subjects receive. Thirdly, excluding the strong stimuli from our analysis amplifies the observed effect. This observation could refer to the repulsive adaptation effect of these strong stimuli (Kohn, 2007). These last two eliminate the effect of the previous stimuli and merely include the decision.

Finally, in order to shed light on the plausible mechanism of the observed effect, we used one successful and elaborate variant of decision making models based on the accumulation-to-bound mechanism called the "drift-diffusion" (Gold & Shadlen, 2007; Kiani et al., 2008; Lerche & Voss, 2017; Mazurek, Roitman, Ditterich, & Shadlen, 2003; Shadlen, Hanks, Churchland, Kiani, & Yang, 2006; Voss, Nagler, & Lerche, 2013; Voss & Voss, 2007). The activity of accumulating information starts from a baseline point toward a criterion level or bound where the decision process is terminated (Bogacz, Brown, Moehlis, Holmes, & Cohen, 2006; Ratcliff, 1978, 2002; Ratcliff, Smith, Brown, & McKoon, 2016). It seems as though, the starting point of evidence accumulation varies depending on the different parameters (Bogacz et al., 2006; Falmagne, 1965, 1968; Forstmann et al., 2010; Luce, 1986; Ratcliff, Van Zandt, & McKoon, 1999; Remington, 1969; Rorie, Gao, McClelland, & Newsome, 2010). Our results show that the model that keeps previous decision information provides a better fit to the behavioral data, leading us to propose that, the state of decision variable (Gold & Shadlen, 2007) after crossing the bound (in the previous decision) can accumulate with decision variable for the current decision in consecutive trials.

## 2. MATERIALS AND METHODS

### 2.1. SUBJECTS

In this experiment, six adult subjects, three males and three females, with normal or correct-to-normal vision participated. All the subjects, except for two, were unfamiliar to the design of the experiment. They signed informed written consent before attending the study. All experimental protocols were approved by the Iran University of Medical Sciences.

### 2.2. VISUAL STIMULI

The stimuli consisted of known dynamic random dot motion used in a verity of perceptual decision making studies. These stimuli are movies in which some dots are randomly moving in different direction. In each frame, white dots (2×2 pixel, 0.088° per side) were displayed on a black background with a density of 16.7 dots/degree2/sec (Roitman & Shadlen, 2002; Shadlen & Newsome, 2001). The stimulus contained three interleaved sets of dots displayed on consecutive video frames. Each set was relocated, three frames (40 ms) later while a fraction of dots had a coherent continuous motion toward a direction and the rest of dots were resettled randomly. The stimulus strength was specified by the fraction of dots which moved coherently. Stimulus was presented using a psychophysics toolbox 3.0.12 (Brainard, 1997; Pelli, 1997) for MATLAB R2013a (The MathWorks Inc, 2013) on a computer with the operating system of windows 7 (64-bit), Intel (R), Core (TM) i7, 16 GB internal storage, and NVIDIA Quadro K2000 GPU card.

### 2.3. BEHAVIORAL TASK

All the experiments were carried out in a semi-dark and sound-proof room. The subjects were seated in an adjustable chair at the distance of 57 cm from a cathode ray tube (CRT) display monitor (19 inch, with an 800×600 screen resolution and 75 Hz refresh rate,). An adjustable chin-rest had been appropriated to support the subject's chin and forehead. Each trial started with a red fixation point (FP, 0.3° diameter) at the center of the screen and two red choice targets (0.5° diameter) on the right and left side of the fixation point (10° eccentricity). The subjects were asked to fix and maintain their gaze on the fixation point throughout the trial. After a 200 ms delay period, the random dots stimulus was displayed within a 5° circular aperture at the center of the screen for 120, 400, and 720 ms. The percentage of coherently moving dots was chosen from these following values: 0%, 3.2%, 6.4%, 12.8%, 25.6%, and 51.2%. At the end of the stimulus presentation, a 120 ms delay period occurred. After the delay period, the Go signal cued the subjects to respond by eliminating the fixation point. The subjects were asked to report their decision, about the direction of motion, within 1 s after the Go signal by pressing a left or right key. Distinctive auditory feedback was delivered for 100 ms for correct responses, error responses and missed trials. The type of feedback was chosen randomly for trials with 0% coherence. Trials have been separated by different gap durations: 0, 120, or 1200 ms (different gap durations were used to demonstrate their different effects on our results, but there was no significant difference between them, so we have pooled the data of the three gaps in all analysis). The arrangement of the motion direction, motion duration, gap duration, and the motion strength varied randomly from trial to trial (Figure 1).

All possible types of trials were randomly interleaved in blocks with 150 trials. The subjects were instructed to perform the experiments quickly and accurately to the possible extent. The overall probability of being correct was shown on the screen at the end of each block. Each subject performed one or two sessions (each session had four blocks) per day until 3600 trials were collected. The results were consistent across all participants but figures have collapsed the data across subjects.

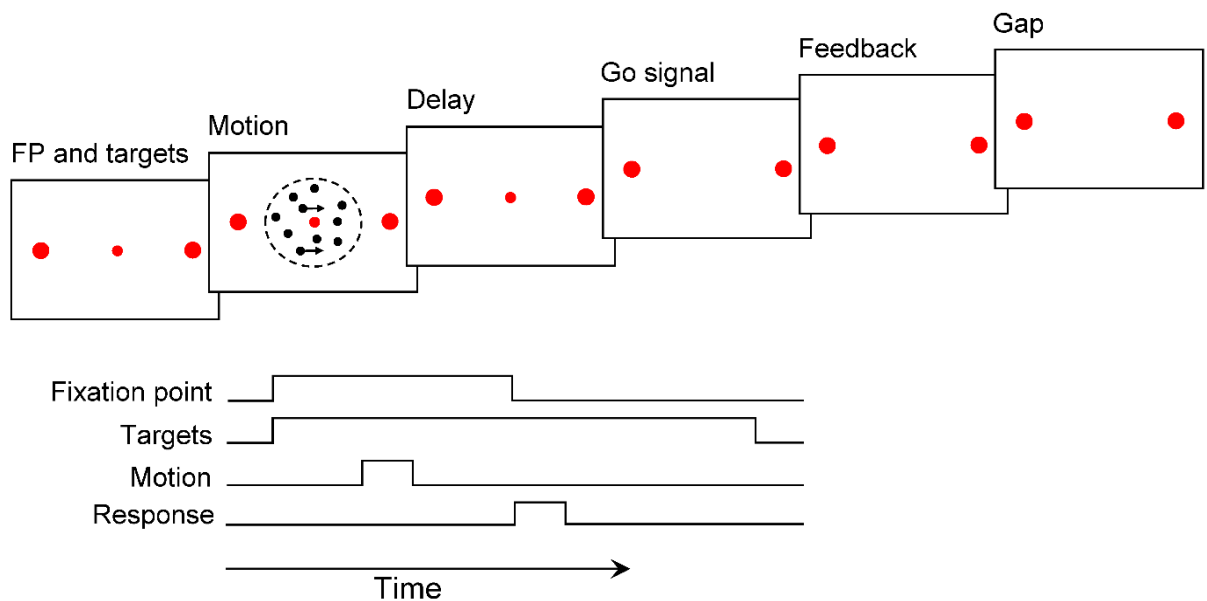

*Figure 1. Motion discrimination paradigm. A fixation point (FP) and two targets were presented for 200 ms. After that, the motion stimulus was shown for 120 ms, 400 ms and 720 ms. The Go signal followed by a 120 ms delay period cued subjects to report their decision, within 1 second, by pressing two specific keys. Auditory feedback was played for 100 ms. The following trial began after a gap of 0-1.2s (See MATERIALS AND METHODS).*

## 2.4. DATA ANALYSIS

For the purpose of this study, we focused our analysis on specific pairs of consecutive trials which will be explained along with their reasons in the following. In order to demonstrate the effect of previous stimulus strengths on the current decision, we picked out only the pair of trials in which the previous trial contains two groups of low (0%, 3.2%) and high (12.8%, 51.2%) motion strengths. In pilot experiments, the results of 25.6%, and 51.2% were similar, so in the main experiment, we chose 12.8% and 51.2% as high motion strengths excerpts in previous trials. We also probed previous trials with three different motion duration to illustrate the effect of previous stimulus durations on the current decision. However, no significant difference was found and since then we have pooled the data of the three motion durations in other analysis. Current trials contained middle motion strength values (3.2%, 6.4%, and 12.8%) and 120 ms motion durations.

A variety of logistic regression models were used to characterize the effect of different parameters on the probability of correct choice. The following models are fitted by using the generalized linear model (GLM) with binomial error structure. We use *Logit[P]* as a short form of $\log(\frac{P}{1-P})$ and $\beta_i$ as fitted coefficients.

The probability of a correct choice is defined by the following (to fit the psychometric function in Figure 2 and for the black curve in Figure 3):

$$\text{Logit}[P_{correct}] = \beta_0 + \beta_1 C \tag{1}$$

where *C* is motion strength. To evaluate the effect of the previous decision on the current choice accuracy, we fit the following: (to fit the psychometric function of same and different decision conditions in Figure 3):

$$\text{Logit}[P_{Correct}] = \beta_0 + \beta_1 I + \beta_2 C, \qquad I = \begin{cases} 0 & \text{different decision} \\ 1 & \text{same decision} \end{cases} \tag{2}$$

where *C* is the motion coherence of the current trials and *I* is an indicator variable for two successive decisions. The null hypothesis is that the current choice accuracy for same and different decision conditions are equal (*H0: $\beta_1=0$*).

A modified version of Equation 2 was used to test whether the current choice accuracy was influenced by correctness and coherence level (low and high) of the previous trial:

$$\text{Logit}[P_{\text{Correct}}] = \beta_0 + \beta_1 I + \beta_2 C + \beta_3 A, \quad (3)$$

$$I = \begin{cases} 0 & \text{different decision} \\ 1 & \text{same decision} \end{cases}, \quad A = \begin{cases} 0 & \text{incorrect (high) previous trial} \\ 1 & \text{correct (low) previous trial} \end{cases}$$

where $C$ is the motion coherence of the current trials and $I$ and $A$ are the indicator variables for two successive decisions and correctness (or coherence level) of the previous trials, respectively. The null hypothesis is that the current choice accuracy does not depend on correctness and coherence level (low and high) of the previous trial ($H0: \beta_3=0$).

To assess the impact of the motion strength of the previous trial on the current choice accuracy we used the following regression:

$$\text{Logit}[P_{\text{Correct}}] = \beta_0 + \beta_1 C_1 + \beta_2 C_2 \quad (4)$$

where $C_1$ and $C_2$ are the motion coherence of the previous and current trials, respectively. The null hypothesis is that previous stimulus strength has no significant effect on current choice accuracy ($H0: \beta_1=0$).

All statistical analyses were performed in R version 3.3.1 (The R Foundation for Statistical Computing, www.R-project.org). The statistical analyses outcomes are presented in the RESULTS section.

### 2.5. MODELING

In order to investigate the mechanism of the last decision impact on the current decision, we used drift-diffusion model (DDM) (Ratcliff, 1978; Ratcliff & McKoon, 2008) implemented by Voss et al. in a computationally efficient, flexible and user-friendly program called fast-dm (Voss & Voss, 2007). Fast-dm estimated DDM's parameters using the partial differential equation (PDE) method through fast computations to calculate the cumulative distribution function (CDF) and the Chi-Square statistic (Lerche & Voss, 2017; Voss et al., 2013).

The diffusion model is undoubtedly a well-established model in the perceptual decision literature (Gold & Shadlen, 2007; Voss et al., 2013). This model is well consistent in both neural and behavioral responses and its different parameters can explain the process of commitment to a choice in the brain based on an accumulation-to-bound mechanism (Gold & Shadlen, 2007; Kiani et al., 2008; Mazurek et al., 2003; Shadlen et al., 2006; Voss & Voss, 2007). In the pure drift-diffusion model (DDM), momentary sensory evidence in favor of one of the choices starts to accumulate from a baseline point $z$. Just after the integrated evidence over time (guided by drift rate $v$) hits a criterion level or bound ($a$), the decision process is terminated (Ratcliff, 1978; Ratcliff & McKoon, 2008; Ratcliff et al., 2016). Seven parameters that exist in the full DDM are divided into three categories: (1) the decision process parameters (decision bound *a*, mean baseline point *z* and mean drift rate *v*), (2) the non-decision process parameter (non-decision time $t_{ND}$), (3) the variability across-trial parameters (variability in stimulus quality $\eta$, variability in baseline point *sz*, and variability in non-decision time $st_{ND}$) (Ratcliff, 1978; Ratcliff & McKoon, 2008; Ratcliff & Tuerlinckx, 2002). According to the proposed hypothesis in the present research, the previous decision can influence the current decision process in three possible ways: (a) the previous decision affects the rate of accumulated evidence (i.e., the drift rate, v) (Ashby, 1983; Bornstein et al., 2017; Diederich & Busemeyer, 2006; Ratcliff, 1985), (b) it changes the mean baseline point of evidence accumulation (*z*) (Bogacz et al., 2006; Bornstein et al., 2017; Diederich & Busemeyer, 2006; Edwards, 1965; Laming, 1968; Link & Heath, 1975; Ratcliff, 1985; Voss, Rothermund, & Voss, 2004; Wagenmakers, Ratcliff, Gomez, & McKoon, 2008) or (c) it shifts the decision threshold (*a*) (Bogacz et al., 2006; Goldfarb et al., 2012; Ratcliff & Rouder, 1998; Ratcliff & Smith, 2004; Simen, Cohen, & Holmes, 2006). The diffusion model will be used to disentangle these three scenarios (Falmagne, 1965, 1968; Luce, 1986; Ratcliff, 1985; Ratcliff & Smith, 2004; Ratcliff et al., 1999; Remington, 1969).

## 3. R E S U L T S
### 3.1. BEHAVIOR

Six human subjects reported the perceived direction of motion in trials with 120, 400 and 720 ms duration (Figure 1). Psychometric function for the subjects is shown in Figure 2. Psychometric function of current trials separated in the three conditions is plotted in Figure 3. The first condition or the so called same decision condition, blue data points, shows the performance of current trials in which the participants have taken a decision similar to the previous trial. In the second condition or different decision condition, red data points, the participants' decisions in current trials are different from those in the previous trials. The third condition, black data points, is the performance of all current trials, independent of the decision in previous trials. Considering the black data points as a reference, an upward and a downward shift is obvious in the psychometric function of the same and different decision conditions, respectively. Shifts in all the three motion strengths were compared by using equivalence test based on Welch's t-test: two one-sided test (TOST) (Berger & Hsu, 1996 ; Gruman, Cribbie, & Arpin-Cribbie, 2007). Applying TOST procedure to test data against equivalence bounds of $d = 0.65$ and $d = -0.65$, revealed that upward shifts in current motion strengths of 3.2% and 6.4% were statistically equivalent, $t(5832.581) = 21.84555$, $p = 0.5 \times 10^{-101}$. The mean difference was $-0.002$, 90% CI [-0.004,-0.001]. As well as upward shifts in current motion strengths of 6.4% and 12.8% were statistically equivalent, $t(5637.74) = 3.332844$, $p = 0.4 \times 10^{-3}$. The mean difference was $-0.015$, 90% CI [-0.016,-0.014]. Therefore, the conclusion from these two TOST was that upward shifts in current motion strengths of 3.2% and 12.8% were statistically equivalent. Furthermore, performing TOST procedure to test data against equivalence bounds of $d = 0.68$ and $d = -0.68$, determined that downward shifts in current motion strengths of 3.2% and 6.4% were statistically equivalent, $t(5826.239) = 25.56059$, $p = 0.6 \times 10^{-136}$. The mean difference was $-0.001$, 90% CI [-0.002,0]. Also downward shifts in current motion strengths of 6.4% and 12.8% were statistically equivalent, $t(5663.589) = 2.587412$, $p = 0.005$. The mean difference was $-0.016$, 90% CI [-0.017,-0.015] (results are summarized in Appendix Figure A1). So, the result from these two TOST was that downward shifts in current motion strengths of 3.2% and 12.8% were statistically equivalent. Hence, in general it can be said that shifts are consistent in all the three motion strengths of current trials in Figure 3.

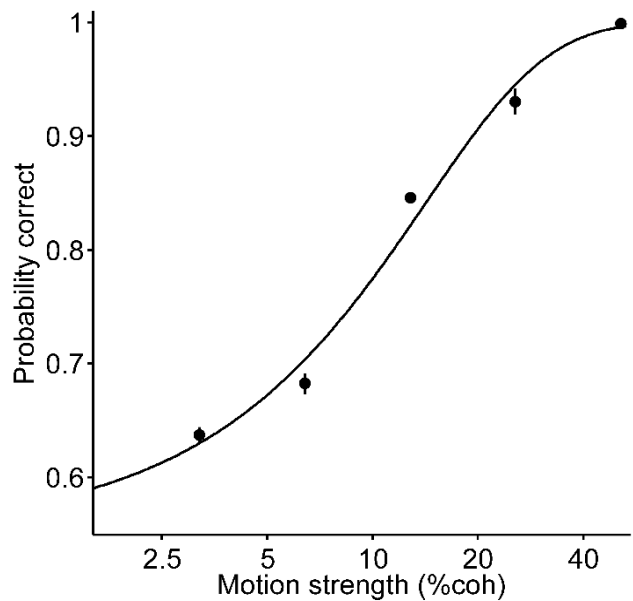

*Figure 2*. Psychometric function of all the trials; each data point presents the performance of pooled data for all the three durations and two directions. The curve is the fit of a logistic regression to the data (Eq. 1). Error bars indicate SE (Standard Error).

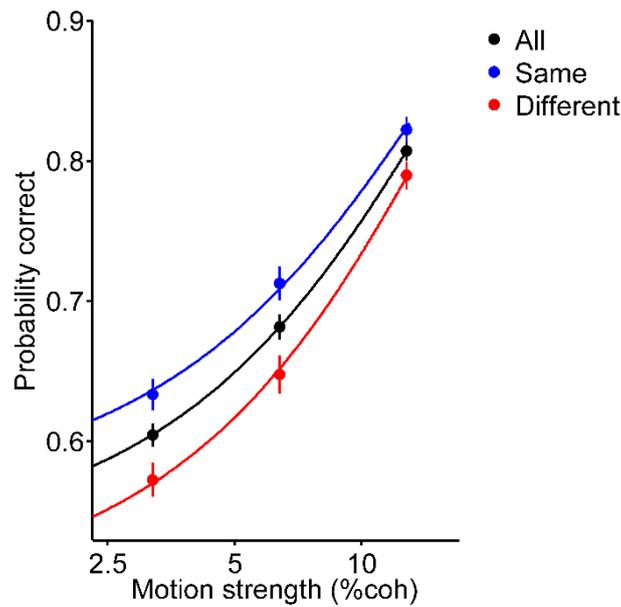

*Figure 3.* Psychometric function of the current trials. Red and blue data points depict performance of subjects for the different and same decision conditions, respectively. Black data points are pooled from these two conditions. Curves are the fit of the logistic regression to the data (Eq. 1 for black curve and Eq. 2 for red and blue curves). Error bars indicate SE (Standard Error).

This difference between psychometric functions of the same and different decision condition implies that the probability of being correct in a decision depends not only on the stimulus strength, but also on the previous decision (Eq. 2, $\beta_1$=0.25±0.09, $p$=5.8×10$^{-8}$). If the stimulus in the current trial has similar (different) direction with the chosen direction in the previous trial, the probability of being correct in this current trial increases (decreases).

Since the previous decision is itself correlated to the previous stimulus, one may conclude that this difference in performance is the effect of stimulus adaptation. Interestingly, the reported effect of the previous decision seems to be in contrast with the repulsive effect of adaptation. Taking the repulsive effect into account, we expect higher sensitivity for the perception of leftward (rightward) motion when it comes after a rightward (leftward) motion. As a result, the probability of being correct should be higher in the different decision condition than in the same one. Below, we tried to elaborate on these two probable contradictory effects through further analysis.

It is worth noting that in case there is an adaptation effect in our paradigm, it should be stronger when the stimulus of the previous trial has high motion coherence. We separated high and low coherence stimuli from the above analysis and calculated how same and different decision conditions differ in performance. Figure 4 illustrates the performance in current trials which includes motion strengths of 3.2%, 6.4% and 12.8%, when previous trials have low motion strengths of 0% and 3.2% (panel A, Eq. 2, $β_1$=0.65±0.13, $p$=4.3×10$^{-22}$) and high motion strengths of 12.8% and 51.2% (panel B, Eq. 2, $β_1$=-0.14±0.13, $p$=0.03). As shown in this figure, the subjects are significantly more probable to choose a correct decision in the different decision condition when the coherence of the previous trial is high, which is consistent with repulsive adaptation effect. Whereas, the panel A in Figure 4 shows that a correct decision is more probable in the same decision condition when stimulus coherence in the previous trial is low (Eq. 3, $β_3$=0.28±0.09, $p$=1.5×10$^{-9}$). The observed effect is significant even when previous trails have 0% coherence. In this coherence, all dots move randomly, which minimizes the adaptation in any specific direction. However, as illustrated in Figure 5, the probability of being correct is greater in the same decision condition than in the different decision condition, even when there is lack of evidence in the previous stimulus (Eq. 2, $β_1$=0.98±0.19, $p$=2.2×10$^{-23}$).

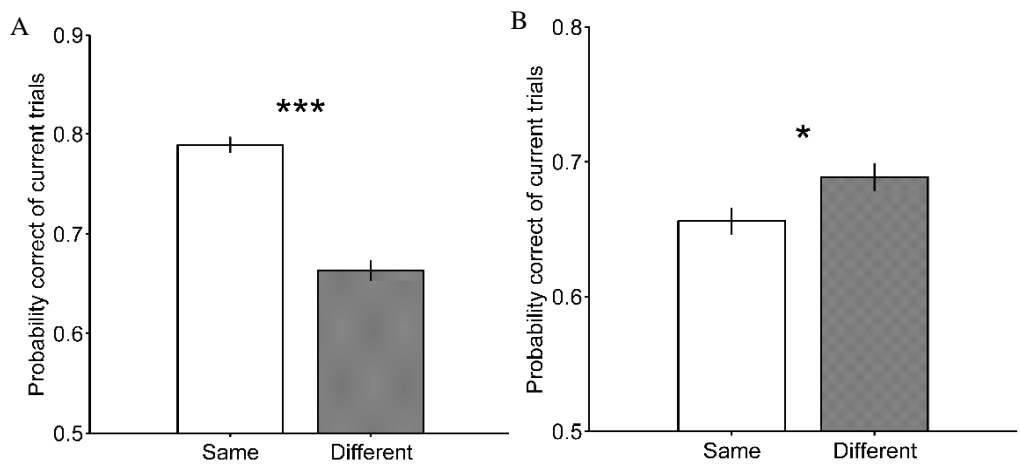

*Figure 4. The performance of the current trials with motion strengths of 3.2%, 6.4%, and 12.8%. Panel A shows performance in the current trials when previous trials have low motion strengths of 0% and 3.2%. Panel B illustrates performance in the current trials when previous trials have high motion strengths of 12.8% and 51.2%. Overall, the same decision condition resulted in greater accuracy than the different decision condition with the low motion strength in the previous trials. Error bars indicate SE (Standard Error). \*$p<0.05$, \*\*\*$p<1E-3$*

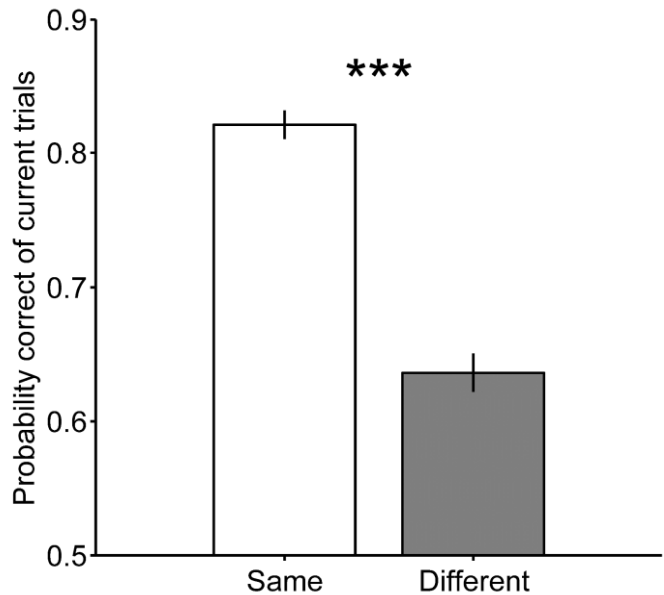

*Figure 5.* *The performance of the current trials which includes motion strengths of 3.2%, 6.4%, and 12.8% when their previous trials are 0%. Error bars indicate SE (Standard Error).* ***$p<1E-3$

As shown in Figure 4, decreasing the effect of stimulus adaptation by excluding current trials with high coherence stimulus seems to strengthen the effect of the previous decision presented in Figure 3. Although results show that probability of being correct in the current decision depends on the previous decision, there is another confounding factor which prevents making any conclusion about the source of this effect. As stated before, feedback signal is different in correct and incorrect trials and this signal may result in the observed effect. Here in Figure 6, by separating correct and incorrect previous trials in both the same and different decision conditions, we attempt to eliminate the effect of feedback. As illustrated in the figure, the correctness of the previous decision does not remove the effect explained above. As shown in Figure 6, similar decision trials are significantly more probable to be correct than different decision trials, regardless of the previous decision to be correct (panel A, Eq. 2, $\beta_1=0.37\pm0.17$, $p=3.3\times10^{-5}$) or incorrect (panel B, Eq. 2, $\beta_1=1.06\pm0.21$, $p=3.6\times10^{-22}$). Indeed, The observed effect in correct and incorrect previous decision conditions are qualitatively (not quantitatively) similar. Therefore, it is only the participant's decision that affects his/her decision in succeeding trials (Eq. 3, $\beta_3=-0.07\pm0.1$, $p=0.19$).

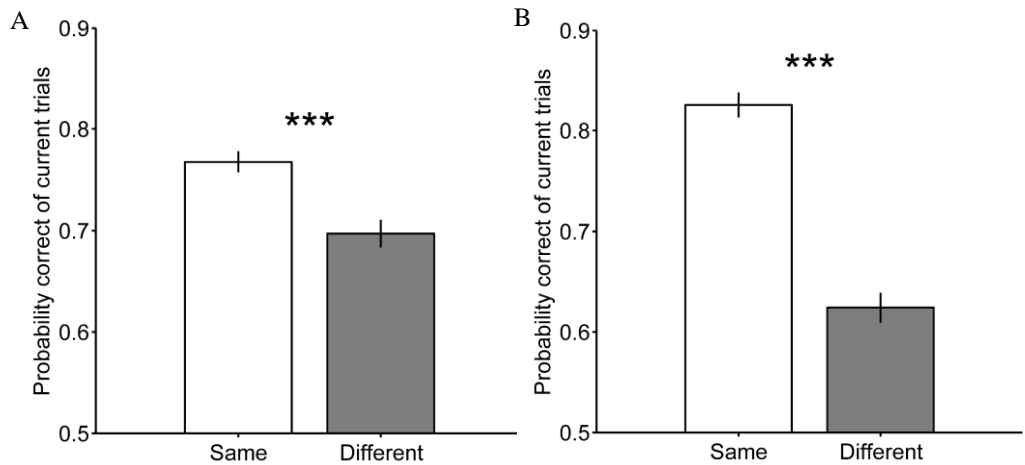

*Figure 6. The performance of the current trials which includes motion strengths of 3.2%, 6.4%, and 12.8%. Panel A is the performance of the current trials when their previous trials are correct with low motion strengths (0% and 3.2%). Panel B is the performance of the current trials when their previous trials are incorrect with low motion strengths (0% and 3.2%). The figure shows that the accuracy of the current trials is higher in the same decision condition compared to the different decision condition for both correct and incorrect previous trials. Error bars indicate SE (Standard Error).\*\*\*p<1E-3*

### 3.2. MODEL FITS

As mentioned above, to investigate the underlying mechanism of the previous decision's effect on the probability of being correct in the current choice, we used the drift-diffusion model (DDM). Dependence of the model parameters on the previous decision gave us the chance to examine the effect of the previous decision on each parameter. To do so, besides of the pure DDM, we ran three modified versions of it and fit these four models to behavioral data derived from experimental study to provide further intuition into the nature of the observed effect.

The first model ($model_p$) is the pure DDM in which the only dependent variable drift-rate ($v$) depends on the current motion strength. In the second model ($model_v$), as a modified DDM, $v$ depends on both current motion strength and previous decision (same and different decision conditions). The third one ($model_z$) is a drift-diffusion model in which the starting point of evidence accumulation ($z$) is dependent on the previous decision, and $v$ depends on the current motion strength. The fourth DDM ($model_a$) modified by the dependence of the decision bound ($a$) on the previous decision, as well as $v$ is dependent on the current motion strength.

Fitted parameters of each model are listed in Tables 1-4 (mean±SE across subjects). For each participant's details, see Appendix Tables A1-A4. Here, $s$ and $d$ indices respectively stand for the same and different decision conditions. As in the second table, based on the dependence of the drift-rate on both current motion strength and previous decision, there are six different drift-rates for three current stimulus coherences (3.2%, 6.4%, and 12.8%) and two conditions (same and different). Regarding the parameters of the third model in Table 3, there are two different starting points for the same and different decision conditions. As presented in Table 4, $model_a$ has two different decision threshold related to two different decision conditions.

As we expected from behavioral results (which indicated the current decision had higher accuracy when the direction of current stimulus was similar to the chosen direction in the previous trial compared to when the direction of current stimulus was different from the chosen direction in the previous trial), when the drift-rate and starting point are dependent on the previous decision, they obtain bigger values in the same decision condition in comparison to the different decision condition and conversely, the decision threshold in the same decision condition is smaller than its value in the different decision condition.

*Table 1. Fitted parameters (mean±SE) of the pure DDM ($model_p$).*

| | |
|---|---|
| $z$ | 0.555±0.015 |
| $a$ | 0.681±0.062 |
| $v_{3.2}$ | 0.321±0.081 |
| $v_{6.4}$ | 0.780±0.113 |
| $v_{12.8}$ | 1.855±0.155 |
| $t_{ND}$ | 0.178±0.008 |
| $st_{ND}$ | 0.116±0.012 |

*Table 2. Fitted parameters (mean±SE) of the second DDM ($model_v$).*

| | |
|---|---|
| $z$ | 0.551±0.013 |
| $a$ | 0.690±0.610 |
| $v_{3.2\_s}$ | 0.488±0.104 |
| $v_{6.4\_s}$ | 1.023±0.136 |
| $v_{12.8\_s}$ | 1.907±0.195 |
| $v_{3.2\_d}$ | 0.201±0.058 |
| $v_{6.4\_d}$ | 0.525±0.100 |
| $v_{12.8\_d}$ | 1.748±0.147 |
| $t_{ND}$ | 0.176±0.008 |
| $st_{ND}$ | 0.120±0.011 |

*Table 3. Fitted parameters* (mean±SE) *of the third DDM (model$_z$).*

|     |             |
| --- | ----------- |
| $z\_s$ | 0.566±0.014 |
| $z\_d$ | 0.540±0.013 |
| $a$   | 0.688±0.061 |
| $v_{3.2}$ | 0.333±0.070 |
| $v_{6.4}$ | 0.755±0.105 |
| $v_{12.8}$ | 1.804±0.135 |
| $t_{ND}$ | 0.176±0.008 |
| $st_{ND}$ | 0.121±0.011 |

*Table 4. Fitted parameters* (mean±SE) *of the fourth DDM (model$_a$).*

|     |             |
| --- | ----------- |
| $z$   | 0.553±0.013 |
| $a\_s$ | 0.685±0.064 |
| $a\_d$ | 0.688±0.059 |
| $v_{3.2}$ | 0.324±0.070 |
| $v_{6.4}$ | 0.758±0.104 |
| $v_{12.8}$ | 1.806±0.131 |
| $t_{ND}$ | 0.176±0.008 |
| $st_{ND}$ | 0.120±0.011 |

Models have compared using the Bayes Information Criterion (BIC) (Kass & Wasserman, 1995; Liddle, 2007; Smith & Spiegelhalter, 1980) for the different model fits are exposed in Table 5 (mean±sd across subjects). For details of subjective scores, see Appendix Tables A5-A8.

As shown in Table 5, the overall quality of the fits was good ($R^2 > 0.83$). Furthermore, the modified DDM with the dependent starting point (model$_z$) received the smallest BIC compared to the model$_p$ ($p=5.6\times10^{-3}$) and model$_a$ ($p=3.2\times10^{-4}$). These significant differences strongly suggested that the model$_z$ is the best one among these three models. Whereas, two subjects' (1 and 2) BIC values in model$_v$ and model$_z$ were close to each other and caused the overall BIC scores were not significantly different in these two models (see Tables A6 and A7). Since these two models yielded nearly equivalent BIC, the simpler model with fewer free parameters confirmed as the preferred model. Eventually, we chose the model$_z$ with the best explanation for how the current choice accuracy is influenced by previous decision.

*Table 5. Model performance comparison via BIC and R2 metrics (mean±sd across subjects).*

| Model | Total parameters | $R^2$ | BIC |
|---|---|---|---|
| $Model_p$ | 7 | 0.836±0.111 | -26.726±6.113 |
| $Model_v$ | 10 | 0.951±0.026 | -29.902±5.570 |
| $Model_z$ | 8 | 0.965±0.016 | -34.858±3.267 |
| $Model_a$ | 8 | 0.843±0.024 | -25.406±1.374 |

In the latest step, we investigated the difference of the dependent parameter in different conditions for the winner model ($model_z$). As stated before, starting point gained higher value in the same decision condition ($z\_s$) compared to the different decision condition ($z\_d$) and it's consistent across all participants except subject 3 (for subjects' details, see Appendix Tables A3). The significance of the differences between $z\_s$ and $z\_d$ was tested by the nonparametric bootstrap method (Efron & Tibshirani, 1994; Hinkley, 1998). These differences were quite significant ($p < 1.7 \times 10^{-165}$) for every five subjects.

## 4. DISCUSSION

Our results showed in sequential perceptual decisions, the probability of being correct in the current choice increases if it is similar to the previous one and conversely decreases when they are different. Against to the views that claim sequential effects (Falmagne, 1965, 1968; Gold et al., 2008; Goldfarb et al., 2012; Remington, 1969) on decision processes are due to the motor response bias or sensory bias (Albright, 2012; Carnevale et al., 2012; Gold et al., 2008; Marcos et al., 2013; Pearson & Brascamp, 2008), this decision history effect cannot be defined through these biases (Akaishi et al., 2014). We hypothesize the state of decision variable, which accumulates information to reach any of the two alternative bounds (Gold & Shadlen, 2007; Kiani et al., 2008), is not reset as soon as making the decision. Even it seems this bound crossing in the previous decision provides information for the subsequent decision and can bias it (Bogacz et al., 2006; Diederich & Busemeyer, 2006; Link & Heath, 1975; Ratcliff, 1985; Voss et al., 2004; Wagenmakers et al., 2008).

To verify this assertion, we presented the results of a behavioral study of decision making using 2AFC paradigm based on randomly moving dots with fixed duration and short interval time, focusing on sequences of two trials. To further study variations in the probability of correct, we extend the pure drift-diffusion model (DDM) (Bogacz et al., 2006; Ratcliff, 1978, 2002; Ratcliff &

Tuerlinckx, 2002) to account for sequential effects (Falmagne, 1965, 1968; Luce, 1986; Ratcliff, 1985; Ratcliff & Smith, 2004; Ratcliff et al., 1999; Remington, 1969). In a modified version of DDM, we proposed a simple mechanism for the dependence of baseline to the previous decision. We indicated how our modified DDM can account for the observed changes in subjects' performance for the same and different decision conditions.

Since in RT tasks which have long Go signals, the time between two decisions in different sequential trials has a lot of fluctuations and might not be short, the sequential effect seems will be diminished. In the current study, to avoid increasing the time between consecutive decisions, we utilized fixed duration task which had fixed period for each part of a trial and limited Go signal. Nevertheless, we recorded the reaction time (time elapsed from Go signal onset to a hand key-press) besides the choice accuracy in our experiment. Although there was no significant difference in reaction times in different decision conditions due to fixed duration task, reaction time decreased with increasing strength of motion (Link, 1992; Ratcliff & Smith, 2004; Roitman & Shadlen, 2002). That's why we used reaction times as the input data of the model in addition to the accuracy, current stimulus strength, and previous decision.

Meanwhile, we did not apply the previous stimulus coherency as the model input considering the behavioral results that provided further support this idea, the beheld sequential effect cannot be caused by the sensory bias. Specifically, $\beta 1$ in Equation 4 is close to zero (Eq. 4, $\beta 1$=-0.003±0.002, $p$=0.017) which implies that the strength of the stimulus in the previous trial does not affect the probability of being correct in the current decision. Though as shown in Figure 4, two different types of motion coherence level (low and high) in the previous trial exert completely opposite effects (decision bias and repulsive adaptation) on the current decision. Indeed, the integration of these two types of previous stimulus coherency in variable $C1$ in Equation 4 led them to neutralize the effect of each other. Similar to what Akaishi et al. (2014) showed the choice repetition probability was significantly higher when low coherence motion was presented on the previous trial than when high coherence motion was presented on the previous trial (Akaishi et al., 2014).

However, our work is principally different from theirs since they indicated the impact of the previous decision on the choice repetition probability (Akaishi et al., 2014) while our purpose is to investigate this effect on the probability of correct in current decision. Another difference is they did not use feedback in their experiments and declared that the mechanism is associated with making an incorrect choice rather than recognition of an error is responsible for the decision bias.

Whereas we claim that the decision, independent of the correctness and having positive or negative feedback, affects the probability of being correct in the next decision (as shown in Figure 6). Moreover, to support this statement we did another analysis by separating correct and incorrect previous trials with 0% motion strength in both the same and different conditions. Actually, we duplicated Figure 6 only for 0% coherent motion of previous trials (See the Appendix Contents, Figure A2). In these trials, all dots had random movements which prevented the sensory bias in any particular direction and feedback was given randomly to the subjects, independently of whether they pressed the left or the right key. So, the subjects received positive feedback on 50% of the trials (Figure A2, panel A) and negative feedback on the other 50% of the trials (Figure A2, panel B). As demonstrated in this figure, similar decision trials are significantly more probable to be correct than different decision trials regardless the kind of previously received feedback.

## ACKNOWLEDGEMENTS

This work was partially supported by the Cognitive Sciences and Technologies Council, Institute for Research in Fundamental Sciences (IPM)-School of Cognitive Sciences (SCS), and Shahid Rajaee Teacher Training University [grant number 29602]. We are thankful to Amirhossein Farzmahdi, Hamed Nili and Abdolhossein Vahabie for helpful discussions.

## CONFLICT OF INTEREST STATEMENT

The authors do not have a conflict of interest.

## AUTHOR CONTRIBUTIONS

F.O has contributed to the conception and study design, data collection, data analysis and interpretation, statistical analysis, modeling and drafting. M.T.M has contributed to the study design, interpretation of data and drafting. S.Z has contributed to the study design, drafting, modeling, data analysis and interpretation. R.E has contributed to the design of the work and interpretation of data. All authors have approved this final version of the manuscript to be published.

# ABBREVIATIONS

ms: millisecond

s: second

sd: standard deviation

BIC: Bayes Information Criterion

CDF: Cumulative Distribution Function

CRT: Cathode Ray Tube

DDM: Drift-Diffusion Model

GLM: Generalized Linear Model

PDE: Partial Differential Equation

$R^2$: R squared

SE: Standard Error

TAFC: Two-Alternative Forced-Choice

TOST: Two One-Sided Test